\documentclass[english,review]{elsarticle}
\usepackage{mathpazo}

\usepackage[T1]{fontenc}
\usepackage[latin9]{inputenc}
\usepackage{geometry}
\usepackage{color}
\usepackage{float}
\usepackage{units}
\usepackage{url}
\usepackage{bm}
\usepackage{amsmath}
\usepackage{amssymb}
\usepackage{cancel}
\usepackage{graphicx}

\makeatletter

\floatstyle{ruled}
\newfloat{algorithm}{tbp}{loa}
\providecommand{\algorithmname}{Algorithm}
\floatname{algorithm}{\protect\algorithmname}

\usepackage{xcolor}

\newcommand{\myBox}[1]{%
  \fbox{$ \displaystyle #1 $}%
}
\usepackage[titletoc]{appendix}

\makeatother

\usepackage{babel}
\begin{document}

\title{A semi-implicit, energy- and charge-conserving particle-in-cell algorithm
for the relativistic Vlasov-Maxwell equations }

\author[]{G. Chen\corref{cor1}}

\ead{gchen@lanl.gov}

\author[]{L. Chacón}

\author[]{L. Yin}

\author[]{B. J. Albright}

\author[]{D. J. Stark}

\author[]{R. F. Bird}

\cortext[cor1]{Corresponding author}

\address{Los Alamos National Laboratory, Los Alamos, NM 87545}
\begin{abstract}
Conventional explicit electromagnetic particle-in-cell (PIC) algorithms
do not conserve discrete energy exactly. Time-centered fully implicit
PIC algorithms can conserve discrete energy exactly, but may introduce
large dispersion errors in the light-wave modes. This can lead to
intolerable simulation errors where accurate light propagation is
needed (e.g. in laser-plasma interactions). In this study, we selectively
combine the leap-frog and Crank-Nicolson methods to produce an exactly
energy- and charge-conserving relativistic electromagnetic PIC algorithm.
Specifically, we employ the leap-frog method for Maxwell's equations,
and the Crank-Nicolson method for the particle equations. The semi-implicit
algorithm admits exact global energy conservation, exact local charge
conservation, and preserves the dispersion properties of the leap-frog
method for the light wave. The algorithm employs a new particle pusher
 designed to maximize efficiency and minimize wall-clock-time impact
vs. the explicit alternative. It has been implemented in a code named
iVPIC, based on the Los Alamos National Laboratory VPIC code (\url{https://github.com/losalamos/vpic}).
We present numerical results that demonstrate the properties of the
scheme with sample test problems: relativistic two-stream instability,
Weibel instability, and  laser-plasma instabilities. 
\end{abstract}
\maketitle

\section{Introduction}

Particle-in-cell (PIC) methods combine Lagrangian and Eulerian techniques
for solving the relativistic Vlasov-Maxwell equations (or a subset
thereof) for kinetic plasma simulations. Specifically, the method
of characteristics is employed to solve Vlasov equation, evolving
macro-particles according to the Newton (or Lorentz) equations of
motion. Finite-difference time-domain (FDTD) methods are commonly
used for evolving Maxwell's equations on a mesh. The Yee scheme \citep{yee1966numerical}
is the most popular algorithm for integrating Maxwell's equations,
due to its simplicity and second-order accuracy. B-splines are widely
used for interpolations between the particles and the mesh. A detailed
description of classical PIC methods and their analysis can be found
in Ref. \citep{birdsall2004plasma,hockney1988computer} .

Despite their numerical simplicity, PIC algorithms have been successfully
applied to a large range of plasma descriptions, including electrostatic
\citep{dawson1962one}, low-frequency electromagnetic (Darwin \citep{hasegawa1968one}),
gyrokinetic \citep{lee1983gyrokinetic}, quasi-static \citep{huang2006quickpic},
and the most general electromagnetic and relativistic systems \citep{langdon1976electromagnetic}.
However, the applicability of PIC for long-term multiscale simulations
may be currently limited by the trustworthiness of the simulation,
as measured by accumulated errors of conservation properties such
as charge, momentum, and energy. Hereafter, we will adopt the notion
of ``discrete conservation'' as conservation satisfied exactly (in
practice to numerical roundoff) in a discrete system (with finite
timestep $\Delta t$ and mesh size $\Delta x$). Discrete charge conservation
is satisfied by most PIC algorithms employed today \citep{villasenor1992rigorous,esirkepov2001exact,chen2011energy,pinto2014charge}.
Discrete energy conservation has been recently demonstrated with advances
in implicit PIC algorithms, either fully implicit \citep{chen2011energy,markidis2011energy,lapenta2011particle,chen2015multi,chacon2016curvilinear},
or semi-implicit \citep{lapenta2017exactly} (note that an early version
of ``energy-conserving'' algorithm by Lewis \citep{lewis1970energy}
conserves discrete energy exactly only in the limit of $\Delta t\rightarrow0$).
Discrete momentum conservation has only been realized in the simplest
electrostatic models integrated by explicit schemes \citep[Ch.8.6]{birdsall2004plasma}.

In this study, we seek to develop an energy- and charge-conserving
scheme for the relativistic Vlasov-Maxwell system. Compared to the
fully implicit approach proposed by Lapenta and Markidis \citep{hasegawa1968one,lapenta2011particle},
which is based on the Crank-Nicolson (CN) scheme, our approach is
less dispersive for light wave modes and consequently less subject
to numerical Cerenkov (or Cherenkov) radiation issues. For the CN
scheme solving Maxwell's equations, the numerical Cerenkov radiation
can be so problematic that, in practice, some numerical dissipation
is needed at the cost of discrete energy conservation \citep{markidis2011energy}.
It is also worth noting that the methods developed by Lapenta and
his group  do not conserve charge (see Ref. \citep{chen2018gauss}
for an improvement on this issue). Energy- and charge-conserving PIC
algorithms based on fully implicit approaches have been developed,
but so far only for electrostatic \citep{chen2011energy} and Vlasov-Darwin
systems (e.g. \citep{chen2015multi}). 

Based on a semi-implicit approach, the proposed energy- and charge-conserving
PIC algorithm has relatively low light-wave dispersion (w.r.t. the
CN scheme) for electromagnetic relativistic plasma simulations. The
algorithm combines the conventional leap-frog scheme for the field
equations with a CN update of the particle equations of motion. The
CN scheme combines the position update of Ref. \citep{lapenta2011particle}
and velocity update of Ref. \citep{higuera2017structure}. Energy
conservation is achieved by implicitly coupling the electric field
advance and particle push, while charge conservation rests on a novel
treatment of particle cell crossings that, unlike previous studies
\citep{chen2011energy,chen2015multi}, demands no particle stopping
at cell boundaries. The leap-frog field update retains a Courant-Friedrics-Lewy
(CFL) time-step stability constraint, which is necessary for stability.
However, respecting the CFL constraint keeps the numerical light-wave
dispersion low, and ensures relatively fast convergence ($\sim5$
Picard iterations) of the electric-field/particle-push coupling. This
approach is mostly motivated by simulations of laser-plasma interactions,
where light wave and relativistic effects dominate. However, the algorithm can
be used for non-relativistic simulations as well.

In what follows, Sec. 2 briefly reviews the classical electromagnetic
relativistic PIC algorithm, and introduces notation. Section 3 introduces
the new algorithmic elements that enable the exact discrete energy
and charge conservation. Section 4 presents some numerical experiments
demonstrating the correctness and long-term conservation properties
of the scheme. We close with some concluding remarks in Sec. 5.

\section{The classical electromagnetic relativistic PIC method}

We consider the relativistic Vlasov-Maxwell system:
\begin{align}
\partial_{t}f_{s}+\mathbf{v}\cdot\nabla f_{s}+\frac{q_{s}}{m_{s}}(\mathbf{E}+\mathbf{v}\times\mathbf{B})\cdot\nabla_{\mathbf{u}}f_{s} & =0,\label{eq:Vlasov}\\
\frac{\partial\mathbf{B}}{\partial t}+\nabla\times\mathbf{E} & =0,\label{eq:Farady's law}\\
\epsilon_{0}\frac{\partial\mathbf{E}}{\partial t}-\frac{1}{\mu_{0}}\nabla\times\mathbf{B}+\mathbf{j} & =0.\label{eq:Ampere's law}\\
\nabla\cdot\mathbf{B} & =0,\label{eq:divB=00003D0}\\
\nabla\cdot\mathbf{E} & =\rho,\label{eq:Gauss's law}
\end{align}
where $f_{s}(\mathbf{x},\mathbf{u})$ is the particle distribution
function of species $s$ in phase space, $\mathbf{x}$ denotes physical
position, $\mathbf{v}$ denotes velocity, $\mathbf{u}=\mathbf{v}\gamma$
is the proper velocity, and $\gamma=\sqrt{1+u^{2}/c^{2}}$ is the
Lorentz factor, with $c$ the speed of light. In these equations,
$\epsilon_{0}$ and $\mu_{0}$ are the vacuum permittivity and permeability
respectively, $\mathbf{E}$ and $\mathbf{B}$ are the electric and
magnetic field, respectively, $q_{s}$ and $m_{s}$ are the species
charge and mass respectively, and $\mathbf{j}$ and $\rho$ are current
and charge densities, respectively, found from the distribution function
as:
\begin{eqnarray*}
\mathbf{j} & = & \sum_{s}q_{s}\int d\mathbf{u}\,\mathbf{v}\,f_{s}(\mathbf{r},\mathbf{u}),\\
\rho & = & \sum_{s}q_{s}\int d\mathbf{u}f_{s}(\mathbf{r},\mathbf{u}).
\end{eqnarray*}
Equations \ref{eq:divB=00003D0} and \ref{eq:Gauss's law} are the
two involutions of Maxwell's equations \citep{seiler2009involution}.
Note that conventional FDTD methods typically advance Eqs. \ref{eq:Farady's law}
and \ref{eq:Ampere's law} only \citep{taflove2005computational}.
Equation \ref{eq:divB=00003D0} is automatically satisfied by Yee's
scheme, and Eq. \ref{eq:Gauss's law} may be enforced by specially
designed divergence-cleaning methods \citep{marder1987method,langdon1992enforcing,munz2000divergence}
when charge conservation 
\begin{equation}
\frac{\partial\rho}{\partial t}+\nabla\cdot\mathbf{j}=0
\end{equation}
is not automatically satisfied. 

Equation \ref{eq:Vlasov} is discretized by samples of $f_{s}(\mathbf{x},\mathbf{u})$
(i.e., macro-particles):
\[
f_{s}(\mathbf{x},\mathbf{u})\approx\sum_{p\in s}w_{p}\delta(\mathbf{x}-\mathbf{x}_{p})\delta(\mathbf{u}-\mathbf{u}_{p}),
\]
where $w_{p}$ is the particle weight (constant in collisionless PIC
simulations). The particle equations of motion read:
\begin{align}
\frac{d\mathbf{x}_{p}}{dt} & =\mathbf{v}_{p},\label{eq:eomx}\\
\frac{d\mathbf{u}_{p}}{dt} & =\frac{q_{p}}{m_{p}}(\mathbf{E}_{p}+\mathbf{v}_{p}\times\mathbf{B}_{p}).\label{eq:eomv}
\end{align}
Assuming the use of a Yee mesh, the classical leap-frog scheme is
written as:
\begin{eqnarray}
\frac{\mathbf{u}_{p}^{n+\nicefrac{1}{2}}-\mathbf{u}_{p}^{n-\nicefrac{1}{2}}}{\Delta t} & = & \frac{q_{p}}{m_{p}}\left(\mathbf{E}_{p}^{n}+\frac{\mathbf{u}_{p}^{n+\nicefrac{1}{2}}+\mathbf{u}_{p}^{n-\nicefrac{1}{2}}}{2\gamma_{p}^{n}}\times\mathbf{B}_{p}^{n}\right),\label{eq:dudt}\\
\frac{\mathbf{x}_{p}^{n+1}-\mathbf{x}_{p}^{n}}{\Delta t} & = & \frac{\mathbf{u}_{p}^{n+\nicefrac{1}{2}}}{\gamma^{n+\nicefrac{1}{2}}},\label{eq:dxdt}\\
\frac{\mathbf{B}_{h}^{n+\nicefrac{1}{2}}-\mathbf{B}_{h}^{n-\nicefrac{1}{2}}}{\Delta t} & = & -\nabla_{h}\times\mathbf{E}_{h}^{n},\label{eq:dBdt}\\
\epsilon_{0}\frac{\mathbf{E}_{h}^{n+1}-\mathbf{E}_{h}^{n}}{\Delta t} & = & \frac{1}{\mu_{0}}\nabla_{h}\times\mathbf{B}_{h}^{n+\nicefrac{1}{2}}-\mathbf{j}_{h}^{n+\nicefrac{1}{2}},\label{eq:dEdt}
\end{eqnarray}
where the superscript $n$ denotes time level, the subscript $p$
denotes a particle quantity or a field evaluated at the particle position,
$\Delta t$ is the timestep, and the subscript $h=(i,j,k)$ denotes
mesh quantities and operators (using Yee finite differences). Here
$\gamma^{n}=\sqrt{1+(\mathbf{u}^{n-\nicefrac{1}{2}}+\frac{q\Delta t}{2m}\mathbf{E}^{n})^{2}}$,
and $\gamma^{n+\nicefrac{1}{2}}=\sqrt{1+(\mathbf{u}^{n+\nicefrac{1}{2}})^{2}}$.
One potential problem with Boris scheme is that it does not preserve
the correct limit in a force-free field~\citep{vay2008simulation}.
The current density $\mathbf{j}_{h}^{n+\nicefrac{1}{2}}$ is gathered
from particles using B-spline interpolation. This scheme features
a Courant-Fredrichs-Lewy (CFL) condition \citep{courant1967partial}
for stability, and constrains cell sizes to be comparable to the Debye
length to suppress finite-grid instabilities \citep{birdsall2004plasma}. 

\section{The energy- and charge-conserving electromagnetic relativistic PIC
method}

We derive a discrete global energy and local charge conserving scheme
for the relativistic Vlasov-Maxwell system by combining the leap-frog
time advance for Maxwell equations and CN for the particle equations.
Energy conservation is achieved by advancing particles and the electric
field synchronously, affording the scheme a semi-implicit character
and demanding iteration. This iteration, however, is simple to implement,
and of rapid convergence. Simultaneously, automatic local discrete
charge conservation is enforced by the proper choice of shape interpolation
functions, and a new treatment of particle cell crossings that does
not require particles to actually stop at cell boundaries, as previously
proposed in Refs. \citep{chen2011energy,chen2015multi}.

\subsection{Conservative semi-implicit Vlasov-Maxwell PIC algorithm}

The proposed algorithm employs the leap-frogged Maxwell update in
Eqs. \ref{eq:dBdt}, \ref{eq:dEdt}, coupled with a time-centered
(CN) particle push (given below). Leap-frog is selected to ensure
relatively low numerical dispersion errors. It is well-known that
the light-wave dispersion relation can be almost perfectly preserved
in 1D with the Yee scheme as $c\Delta t/\Delta x\rightarrow1$. In
multiple dimensions, the numerical dispersion varies with modal wavelength,
propagation direction, and spatial discretization \citep{taflove2005computational},
but it remains tolerable in practice for many applications. In contrast,
for arbitrary $\Delta t$, the CN scheme applied to Maxwell equations
\citep{sun2003unconditionally,sun2004unconditionally} distorts the
light wave dispersion relation, and slows down the light wave phase
speed significantly as $k\Delta x\rightarrow\pi$ \citep{chen2014energy}.
Charged particles with speed greater than the phase speed of light
waves would lose much of their energy via Cerenkov radiation \citep{jackson2012classical},
and may generate significant noise in the simulation. For this reason,
we keep leap-frog for the field time advance. For the analysis below,
we rewrite Eq. \ref{eq:dBdt} in the following equivalent way:
\begin{align}
\frac{\mathbf{B}_{h}^{n+\nicefrac{1}{2}}-\mathbf{B}_{h}^{n}}{\Delta t/2} & =-\nabla_{h}\times\mathbf{E}_{h}^{n},\label{eq:dBdt-1}\\
\frac{\mathbf{B}_{h}^{n+1}-\mathbf{B}_{h}^{n+\nicefrac{1}{2}}}{\Delta t/2} & =-\nabla_{h}\times\mathbf{E}_{h}^{n+1}.\label{eq:dBdt-2}
\end{align}

Particles are advanced fully implicitly by the CN scheme, yielding:
\begin{align}
\frac{\mathbf{x}_{p}^{n+1}-\mathbf{x}_{p}^{n}}{\Delta t} & =\mathbf{v}_{p}^{n+\nicefrac{1}{2}},\label{eq:CN-x}\\
\frac{\mathbf{u}_{p}^{n+1}-\mathbf{u}_{p}^{n}}{\Delta t} & =\frac{q_{p}}{m_{p}}(\mathbf{E}_{p}^{n+\nicefrac{1}{2}}+\bar{\mathbf{v}}_{p}^{n+\nicefrac{1}{2}}\times\mathbf{B}_{p}^{n+\nicefrac{1}{2}}),\label{eq:CN-v}
\end{align}
where we define 
\begin{eqnarray}
\mathbf{v}_{p}^{n+\nicefrac{1}{2}} & = & \frac{\mathbf{u}_{p}^{n+1}+\mathbf{u}_{p}^{n}}{\gamma_{p}^{n+1}+\gamma_{p}^{n}},\label{eq:v_half}\\
\bar{\mathbf{v}}_{p}^{n+\nicefrac{1}{2}} & = & \frac{\mathbf{u}_{p}^{n+1}+\mathbf{u}_{p}^{n}}{2\sqrt{1+\left(\frac{\mathbf{u}_{p}^{n+1}+\mathbf{u}_{p}^{n}}{2c}\right)^{2}}}.\label{eq:vbar_half}
\end{eqnarray}
Note that this particle pusher is similar to that in Ref. \citep{lapenta2011particle}
in using $\mathbf{v}_{p}^{n+\nicefrac{1}{2}}$ in the particle position
update, but it employs $\bar{\mathbf{v}}_{p}^{n+\nicefrac{1}{2}}$
for velocity the update, as in Ref. \citep{higuera2017structure}.
This push can be made energy- and charge-conserving, as shown in the
following sections. The field interpolations are given by:
\begin{align}
\mathbf{E}_{p}^{n+\nicefrac{1}{2}} & =\sum_{h}\mathbf{E}_{h}^{n+\nicefrac{1}{2}}\cdot\bar{\bar{\mathbf{S}}}(\mathbf{x}_{h}-\mathbf{x}_{p}),\label{eq:Ep^n+1/2}\\
\mathbf{B}_{p}^{n+\nicefrac{1}{2}} & =\sum_{h}\mathbf{B}_{h}^{n+\nicefrac{1}{2}}\cdot\bar{\bar{\mathbf{S}}}(\mathbf{x}_{h}-\mathbf{x}_{p}).\label{eq:Bp^n+1/2}
\end{align}
Note that the field $\mathbf{E}_{h}$ and $\mathbf{B}_{h}$ at half
timestep are obtained differently: 
\begin{equation}
\mathbf{E}_{h}^{n+\nicefrac{1}{2}}=\frac{\mathbf{E}_{h}^{n+1}+\mathbf{E}_{h}^{n}}{2}\label{eq:Ehalf}
\end{equation}
and $\mathbf{B}_{h}^{n+\nicefrac{1}{2}}$ is obtained from Eq. \ref{eq:dBdt-1}.
Here, $\mathbf{x}_{h}$ is the grid location, $\mathbf{x}_{p}$ is
typically chosen to be at the center of particle trajectory \citep{chen2011energy},
and as before the subscript $h=(i,j,k)$ denotes the grid index. The
specific form of the shape function dyad $\bar{\bar{\mathbf{S}}}$
and Eq. \ref{eq:v_half} are important to ensure energy and charge
conservation, and will be discussed in Secs. \ref{subsec:Energy-conservation}
and \ref{subsec:Charge-conservation}. We employ a direct inversion
of Eq. \ref{eq:CN-v}, as described in Ref. \citep{higuera2017structure}.

The conservative Vlasov-Maxwell PIC algorithm is closed with the following
definition for the current density:
\begin{equation}
\mathbf{j}_{h}^{n+\nicefrac{1}{2}}=\frac{1}{\bm{\Delta}_{h}}\sum_{p}q_{p}\mathbf{v}_{p}^{n+\nicefrac{1}{2}}\cdot\bar{\bar{\mathbf{S}}}(\mathbf{x}_{h}-\mathbf{x}_{p}).\label{eq:charge-deposition}
\end{equation}
where $\bm{\Delta}_{h}$ is the cell volume. Note that, we have used
identical shape functions for the electric field (Eq. \ref{eq:Ep^n+1/2})
and the current density (Eq. \ref{eq:charge-deposition}) to ensure
exact energy conservation \citep{chen2011energy,chen2015multi}. Also
note that time-centered update of particle positions and velocities,
together with electric field, results in a coupled field-particle
system: $\mathbf{x}_{p}$ is a function of the new-time electric field
through Eqs. \ref{eq:CN-x} and \ref{eq:CN-v}, which in turn determines
the current that determines the field via Eq. \ref{eq:dEdt}. This,
in turn, will require an iterative solve, which will be introduced
later (Sec. \ref{subsec:Iterative-algorithm}).

We prove the energy conservation theorem next. 

\subsection{Energy conservation\label{subsec:Energy-conservation}}

Discrete energy conservation can be readily shown as follows. Multiplying
Eq. \ref{eq:dEdt} by $\mathbf{E}_{h}^{n+\nicefrac{1}{2}}$ and integrating
over space, we find: 
\begin{equation}
\sum_{h}\bm{\Delta}_{h}\left[\epsilon_{0}(\mathbf{E}_{h}^{n+1}-\mathbf{E}_{h}^{n})\cdot\mathbf{E}_{h}^{n+\nicefrac{1}{2}}-\frac{\Delta t}{\mu_{0}}(\nabla_{h}\times\mathbf{B}_{h}^{n+\nicefrac{1}{2}})\cdot\mathbf{E}_{h}^{n+\nicefrac{1}{2}}+\Delta t\mathbf{j}_{h}^{n+\nicefrac{1}{2}}\cdot\mathbf{E}_{h}^{n+\nicefrac{1}{2}}\right]=0.\label{eq:EdEdt}
\end{equation}
The first term in Eq. \ref{eq:EdEdt} gives the change of electric
energy: 
\[
\frac{\epsilon_{0}}{2}\sum_{h}\bm{\Delta}_{h}(\mathbf{E}_{h}^{n+1}-\mathbf{E}_{h}^{n})\cdot\mathbf{E}_{h}^{n+\nicefrac{1}{2}}=\frac{\epsilon_{0}}{2}\sum_{h}\bm{\Delta}_{h}\left[(\mathbf{E}_{h}^{n+1})^{2}-(\mathbf{E}_{h}^{n})^{2}\right]\equiv W_{E}^{n+1}-W_{E}^{n},
\]
using Eq. \ref{eq:Ehalf}. The second term gives the change of magnetic
energy: 
\begin{eqnarray*}
-\frac{\Delta t}{\mu_{0}}\sum_{h}\bm{\Delta}_{h}(\nabla_{h}\times\mathbf{B}_{h}^{n+\nicefrac{1}{2}})\cdot\mathbf{E}_{h}^{n+\nicefrac{1}{2}} & = & -\frac{\Delta t}{\mu_{0}}\sum_{h}\bm{\Delta}_{h}(\nabla_{h}\times\frac{\mathbf{E}_{h}^{n+1}+\mathbf{E}_{h}^{n}}{2})\cdot\mathbf{B}_{h}^{n+\nicefrac{1}{2}}\\
 & = & \frac{1}{\mu_{0}}\sum_{h}\bm{\Delta}_{h}\left[(\mathbf{B}_{h}^{n+1}-\mathbf{B}_{h}^{n})\cdot\mathbf{B}_{h}^{n+\nicefrac{1}{2}}\right]=W_{B}^{n+1}-W_{B}^{n},
\end{eqnarray*}
where we have used discrete integration by parts, Eqs. \ref{eq:dBdt-2}
and \ref{eq:dBdt-1}, and we have defined the magnetic energy as:
\begin{equation}
W_{B}^{n}\equiv\frac{1}{2\mu_{0}}\sum_{h}\bm{\Delta}_{h}\mathbf{B}_{h}^{n+\nicefrac{1}{2}}\cdot\mathbf{B}_{h}^{n-\nicefrac{1}{2}},\label{eq:magnetic energy}
\end{equation}
using that $\mathbf{B}_{h}^{n}=(\mathbf{B}_{h}^{n+\nicefrac{1}{2}}+\mathbf{B}_{h}^{n-\nicefrac{1}{2}})/2$
(see Eq. \ref{eq:dBdt-1}). This definition is non-standard, and as
we show in Appendix. \ref{sec:Well-posedness-of-magnetic}, it is
almost always well posed as long as the CFL condition is respected.

The last term in Eq. \ref{eq:EdEdt} equals the change in kinetic
energy: 
\begin{eqnarray*}
\Delta t\sum_{h}\bm{\Delta}_{h}\mathbf{j}_{h}^{n+\nicefrac{1}{2}}\cdot\mathbf{E}_{h}^{n+\nicefrac{1}{2}} & = & \Delta t\sum_{h}\mathbf{E}_{h}^{n+\nicefrac{1}{2}}\cdot\sum_{p}q_{p}\mathbf{v}_{p}^{n+\nicefrac{1}{2}}S(\mathbf{x}_{p}-\mathbf{x}_{h})=\sum_{p}m_{p}\mathbf{v}_{p}^{n+\nicefrac{1}{2}}\cdot(\mathbf{u}_{p}^{n+1}-\mathbf{u}_{p}^{n})\\
 & = & \sum_{p}m_{p}c^{2}(\gamma_{p}^{n+1}-\gamma_{p}^{n})\equiv W_{p}^{n+1}-W_{p}^{n},
\end{eqnarray*}
where we have used Eqs. \ref{eq:CN-v}, \ref{eq:v_half}, \ref{eq:vbar_half},
and that 
\begin{equation}
\mathbf{v}_{p}^{n+\nicefrac{1}{2}}\cdot\bar{\mathbf{v}}_{p}^{n+\nicefrac{1}{2}}\times\mathbf{B}_{p}^{n+\nicefrac{1}{2}}=0.\label{eq:Lorentz_work}
\end{equation}
The energy conservation theorem sought follows:
\[
\left.\left(W_{E}+W_{B}+W_{p}\right)\right|_{n}^{n+1}=0.
\]

\subsection{Charge conservation\label{subsec:Charge-conservation}}

Discrete local charge conservation requires $\partial_{t}\rho+\nabla\cdot\mathbf{j}=0$
at every grid point $(i,j,k)$. Because the charge conservation equation
is linear, it is sufficient to enforce this constraint on the contributions
of each particle:
\begin{equation}
\frac{(\rho_{p})_{ijk}^{n+1}-(\rho_{p})_{ijk}^{n}}{\Delta t}+\left.\nabla_{h}\cdot\mathbf{j}_{p}^{n+\nicefrac{1}{2}}\right|_{ijk}=0.\label{eq:disc-cc}
\end{equation}
We employ first-order (trilinear) splines for the charge density:
\begin{equation}
\rho_{p,ijk}^{n}=q_{p}S_{1}(x_{i}-x_{p}^{n})S_{1}(y_{j}-y_{p}^{n})S_{1}(z_{k}-z_{p}^{n}),\label{eq:rhon}
\end{equation}
and mixed zeroth- and first-order splines (indicated by subscripts
0 and 1 respectively) for the current density: 
\begin{eqnarray}
j_{x,p,i+\nicefrac{1}{2},j,k}^{n+\nicefrac{1}{2}} & = & q_{p}v_{x}^{n+\nicefrac{1}{2}}S_{0}(x_{i+\nicefrac{1}{2}}-x_{p}^{n+\nicefrac{1}{2}})\mathcal{\mathbb{S}}_{11,jk}^{n+\nicefrac{1}{2}}(y_{p},z_{p}),\label{eq:jx}\\
j_{y,p,i,j+\nicefrac{1}{2},k}^{n+\nicefrac{1}{2}} & = & q_{p}v_{y}^{n+\nicefrac{1}{2}}S_{0}(y_{j+\nicefrac{1}{2}}-y_{p}^{n+\nicefrac{1}{2}})\mathcal{\mathbb{S}}_{11,ik}^{n+\nicefrac{1}{2}}(z_{p},x_{p}),\label{eq:jy}\\
j_{z,p,i,j,k+\nicefrac{1}{2}}^{n+\nicefrac{1}{2}} & = & q_{p}v_{z}^{n+\nicefrac{1}{2}}S_{0}(z_{k+\nicefrac{1}{2}}-z_{p}^{n+\nicefrac{1}{2}})\mathcal{\mathbb{S}}_{11,ij}^{n+\nicefrac{1}{2}}(x_{p},y_{p}),\label{eq:jz}
\end{eqnarray}
where, for instance 
\begin{align*}
\mathcal{\mathbb{S}}_{11,jk}^{n+\nicefrac{1}{2}}(y_{p},z_{p})\equiv\frac{1}{3} & \left[S_{1}(y_{j}-y_{p}^{n+1})S_{1}(z_{k}-z_{p}^{n+1})+\frac{S_{1}(y_{j}-y_{p}^{n})S_{1}(z_{k}-z_{p}^{n+1})}{2}\right.\\
 & \left.\:\;+\frac{S_{1}(y_{j}-y_{p}^{n+1})S_{1}(z_{k}-z_{p}^{n})}{2}+S_{1}(y_{j}-y_{p}^{n})S_{1}(z_{k}-z_{p}^{n})\right],
\end{align*}
and so on. The proof of charge conservation is as follows. We first
decompose the density change into three one-dimensional shifts (letting
$q_{p}=1$): 
\begin{eqnarray*}
\rho_{ijk}^{n+1}-\rho_{ijk}^{n} & = & S_{1}(x_{i}-x_{p}^{n+1})S_{1}(y_{j}-y_{p}^{n+1})S_{1}(z_{k}-z_{p}^{n+1})-S_{1}(x_{i}-x_{p}^{n})S_{1}(y_{j}-y_{p}^{n})S_{1}(z_{k}-z_{p}^{n})\\
 & = & \left[S_{1}(x_{i}-x_{p}^{n+1})-S_{1}(x_{i}-x_{p}^{n})\right]\mathcal{\mathbb{S}}_{11,jk}^{n+\nicefrac{1}{2}}(y_{p},z_{p})+\\
 &  & \left[S_{1}(y_{j}-y_{p}^{n+1})-S_{1}(y_{j}-y_{p}^{n})\right]\mathbb{S}_{11,ik}^{n+\nicefrac{1}{2}}(z_{p},x_{p})+\\
 &  & \left[S_{1}(z_{k}-z_{p}^{n+1})-S_{1}(z_{k}-z_{p}^{n})\right]\mathbb{S}_{11,ij}^{n+\nicefrac{1}{2}}(x_{p},y_{p}).
\end{eqnarray*}
Along each direction, using the particle orbit position equation,
it can be shown that (see Appendix \ref{sec:Equvilence-to-Villasenor}
and Ref. \citep{chen2011energy}):
\begin{equation}
\frac{S_{1}(x_{i}-x_{p}^{n+1})-S_{1}(x_{i}-x_{p}^{n})}{\Delta t}+v_{x}^{n+\nicefrac{1}{2}}\frac{S_{0}(x_{i+\nicefrac{1}{2}}-x_{p}^{n+\nicefrac{1}{2}})-S_{0}(x_{i-\nicefrac{1}{2}}-x_{p}^{n+\nicefrac{1}{2}})}{\Delta x}=0,\label{eq:cc-1d}
\end{equation}
where 
\begin{equation}
x_{p}^{n+\nicefrac{1}{2}}=\frac{x_{p}^{n+1}+x_{p}^{n}}{2},\label{eq:xp_centered}
\end{equation}
and similarly for the$\hat{y}$ and $\hat{z}$ directions. Equation
\ref{eq:disc-cc} then follows. The proof requires that the particle
trajectory be within a cell (without crossing a cell boundary), but
is generalized in the next section. It is worth noting that the derivation
shown here results exactly the same charge and current deposition
scheme as Villasenor \& Buneman's \citep{villasenor1992rigorous}
(See Appendix. \ref{sec:Equvilence-to-Villasenor}). Cell crossings
are generalized to ensure simultaneous charge and energy conservation
(Sec. \ref{subsec:Particle-cell-crossing}). Moreover, the derivation
can be extended to second-order shape functions for charge density
\citep{stanier2019fully}. 

From the previous discussion, it is now clear that the shape function
dyad introduced in Eqs. \ref{eq:Ep^n+1/2}, \ref{eq:Bp^n+1/2} and
\ref{eq:charge-deposition} must be defined as:
\begin{eqnarray*}
\bar{\bar{\mathbf{S}}}(\mathbf{x}_{ijk}-\mathbf{x}_{p}) & = & \mathbf{i}\otimes\mathbf{i}\,\,S_{0}(x_{i}-x_{p})\mathcal{\mathbb{S}}_{11,jk}^{n+\nicefrac{1}{2}}(y_{p},z_{p})\\
 & + & \mathbf{j}\otimes\mathbf{j}\,\,S_{0}(y_{j}-y_{p})\mathcal{\mathbb{S}}_{11,ik}^{n+\nicefrac{1}{2}}(z_{p},x_{p})\\
 & + & \mathbf{k}\otimes\mathbf{k}\,\,S_{0}(z_{k}-z_{p})\mathcal{\mathbb{S}}_{11,ij}^{n+\nicefrac{1}{2}}(x_{p},y_{p}),
\end{eqnarray*}
where $\mathbf{i}$, $\mathbf{j}$, $\mathbf{k}$ are Cartesian unit
vectors. 

\subsection{Particle cell crossing\label{subsec:Particle-cell-crossing}}

To ensure discrete charge and energy conservation, previous studies
\citep{chen2011energy,chen2015multi} have advocated employing particle
subcycling to deal with cell crossing. In this approach, each particle
substep is performed with a CN step, which requires a nonlinear solve
(e.g., by Picard iteration). While effective, this subcycling strategy
can be expensive when multiple crossings occur in a single timestep
(e.g., at cell corners), or when trajactory turning points occur near
cell faces and the nonlinear iteration converges slowly. 

Here, we introduce a simpler approach to avoid subcycling at cell
crossings, which is critical for the overall efficiency of the algorithm.
We begin with the assumption that the trajectory is a straight line
during the (CFL-constrained) timestep $\Delta t$ (in this aspect,
it is similar to that used in Ref. \citep{villasenor1992rigorous}).
Thus, we rewrite Eq. \ref{eq:CN-x} and \ref{eq:CN-v} as: 
\begin{align}
\mathbf{x}_{p}^{n+1}-\mathbf{x}_{p}^{n} & =\mathbf{v}_{p}^{n+\nicefrac{1}{2}}\Delta t,\label{eq:CN-x-1}\\
\mathbf{u}_{p}^{n+1}-\mathbf{u}_{p}^{n} & =\sum_{\nu=0}^{N_{\nu}-1}\mathbf{a}_{p}^{\nu+\nicefrac{1}{2}}\Delta\tau_{p}^{\nu},\label{eq:CN-v-1}
\end{align}
where the superscript $\nu$ denotes a sub-step. Here a sub-step is
defined as a trajectory segment within a cell, and $\mathbf{a}_{p}^{\nu+\nicefrac{1}{2}}=\frac{q_{p}}{m_{p}}\sum_{h}\mathbf{E}_{h}^{n+\nicefrac{1}{2}}\cdot\bar{\bar{\mathbf{S}}}(\mathbf{x}_{h}-\mathbf{x}_{p}^{\nu+\nicefrac{1}{2}})$
is the acceleration during that segment of sub-step $\nu$ define
by $[\mathbf{x}_{p}^{\nu},\mathbf{x}_{p}^{\nu+1}]$, evaluated at
its center, 
\begin{equation}
\mathbf{x}_{p}^{\nu+\nicefrac{1}{2}}=\frac{\mathbf{x}_{p}^{\nu+1}+\mathbf{x}_{p}^{\nu}}{2}.\label{eq:xp_centered-1}
\end{equation}
The number of sub-steps $N_{\nu}$ is determined by the number of
cell-crossings. If we define the trajectory segment to be $\Delta\mathbf{x}_{p}^{\nu}=\mathbf{x}_{p}^{\nu+1}-\mathbf{x}_{p}^{\nu}=\mathbf{v}_{p}^{n+\nicefrac{1}{2}}\Delta\tau^{\nu}$,
then the sub-timestep $\Delta\tau^{\nu}$ is given by: 
\[
\Delta\tau_{p}^{\nu}=\frac{|\Delta\mathbf{x}_{p}^{\nu}|}{|\Delta\mathbf{x}_{p}|}\Delta t,
\]
where $\Delta\mathbf{x}_{p}=\mathbf{x}_{p}^{n+1}-\mathbf{x}_{p}^{n}$.

The definitions in Eqs. \ref{eq:CN-x-1}, \ref{eq:CN-v-1} lead straightforwardly
to discrete charge and energy conservation, as before \citep{chen2011energy,chen2015multi},
with an orbit-averaged current density along the trajectory as:
\begin{equation}
\bar{\mathbf{j}}_{h}=\frac{1}{\bm{\Delta}_{h}\Delta t}\sum_{p}q_{p}\sum_{\nu=0}^{N_{\nu}-1}\Delta\mathbf{x}_{p}^{\nu}\cdot\bar{\bar{\mathbf{S}}}(\mathbf{x}_{h}-\mathbf{x}_{p}^{\nu+\nicefrac{1}{2}}).\label{eq:j_avg}
\end{equation}
It is worth pointing out the key conditions for discrete energy and
charge conservation are:
\begin{enumerate}
\item Identical shape functions are used for current deposition (Eq. \ref{eq:j_avg})
and electric field interpolation (Eq. \ref{eq:Ep^n+1/2});
\item The Lorentz force does no work on particles (Eq. \ref{eq:Lorentz_work}). 
\item The shape functions are evaluated at the center of each trajectory
segment (Eq. \ref{eq:xp_centered-1}).
\end{enumerate}

\subsection{Preservation of the Maxwell involutions\label{subsec:Preservation-of-the}}

As stated earlier, Maxwell's equations feature two involutions: the
solenoidal character of the magnetic field and Gauss' law. Regarding
the former, we recall that the standard Yee scheme for Maxwell's equations
enforces the solenoidal property of magnetic field discretely, i.e.,
\begin{equation}
\nabla_{h}\cdot\mathbf{B}=0,\label{eq:divergence-free-B}
\end{equation}
as long as the initial magnetic field is divergence free. This is
seen by taking the discrete divergence of Eq. \ref{eq:dBdt}, and
noting that 
\begin{equation}
\nabla_{h}\cdot\nabla_{h}\times\mathbf{B}=0,\label{eq:div_curl}
\end{equation}
discretely by the Yee scheme. In practice, the solenoidal property
is satisfied to numerical round-off levels.

Gauss' law is enforced in our scheme by the exact charge conservation
property of the particle update, as follows. Taking the divergence
of Ampere's law (Eq. \ref{eq:dEdt}) and utilizing Eq. \ref{eq:div_curl},
there results:
\[
\epsilon_{0}\frac{\nabla_{h}\cdot\mathbf{E}_{h}^{n+1}-\nabla_{h}\cdot\mathbf{E}_{h}^{n}}{\Delta t}+\nabla_{h}\cdot\mathbf{j}_{h}^{n+\nicefrac{1}{2}}=0.
\]
The conservation of charge independently enforces:
\[
\nabla_{h}\cdot\mathbf{j}_{h}^{n+\nicefrac{1}{2}}=-\frac{\rho_{h}^{n+1}-\rho_{h}^{n}}{\Delta t}.
\]
Combining these two equations, we find:
\[
\nabla_{h}\cdot\mathbf{E}_{h}^{n+1}-\nabla_{h}\cdot\mathbf{E}_{h}^{n}=\rho_{h}^{n+1}-\rho_{h}^{n},
\]
which implies Gauss' law at every timestep provided that it be satisfied
in the beginning.

\subsection{Iterative algorithm\label{subsec:Iterative-algorithm}}

\begin{algorithm}
\caption{\label{alg:Iterative-solution}Iterative solution of the semi-implicit
conservative Vlasov-Maxwell PIC equations}

\begin{enumerate}
\item Start from state $\mathbf{B}_{h}^{n-\nicefrac{1}{2}}$, $\mathbf{E}_{h}^{n}$,
and $\left\{ \mathbf{x}_{p}^{n},\mathbf{v}_{p}^{n}\right\} $.
\item Update magnetic field to $\mathbf{B}_{h}^{n+\nicefrac{1}{2}}$ according
to Eq. \ref{eq:dBdt}.
\item Solve the coupled Ampere-particle system (Eqs. \ref{eq:dEdt}, \ref{eq:CN-x},
\ref{eq:CN-v}) for $\mathbf{E}_{h}^{n+1}$, $\mathbf{x}_{p}^{n+1}$,
and $\mathbf{v}_{p}^{n+1}$ by the following iterative procedure ($k$
is the iteration index; lagged electric field coupling is indicated
with a box).
\begin{eqnarray*}
\mathrm{Starting\:with\:\mathbf{E}_{h}^{n+1,k=0}} & = & \mathbf{E}_{h}^{n}.\;\mathrm{For\:each\:iteration}:\\
\mathbf{x}_{p}^{n+1,k}-\mathbf{x}_{p}^{n} & = & \mathbf{v}_{p}^{n+\nicefrac{1}{2},k}\Delta t,\\
\mathbf{u}_{p}^{n+1,k}-\mathbf{u}_{p}^{n} & = & \sum_{\nu}\frac{q_{p}}{m_{p}}\left(\frac{\myBox{\mathbf{E}_{p}^{n+1,k}}+\mathbf{E}_{p}^{n}}{2}+\bar{\mathbf{v}}_{p}^{n+\nicefrac{1}{2},k}\times\mathbf{B}_{p}^{n+\nicefrac{1}{2}}\right)\Delta\tau^{\nu},\\
\bar{\mathbf{j}}_{h} & = & \frac{1}{\bm{\Delta}_{h}\Delta t}\sum_{p}q_{p}\mathbf{v}_{p}^{n+\nicefrac{1}{2},k}\cdot\sum_{\nu}\bar{\bar{\mathbf{S}}}(\mathbf{x}_{h}-\mathbf{x}_{p}^{\nu+\nicefrac{1}{2}})\Delta\tau^{\nu},\\
\epsilon_{0}\frac{\myBox{\mathbf{E}_{h}^{n+1,k+1}}-\mathbf{E}_{h}^{n}}{\Delta t} & = & \frac{1}{\mu_{0}}\nabla_{h}\times\mathbf{B}_{h}^{n+\nicefrac{1}{2}}-\bar{\mathbf{j}}_{h},\\
\mathrm{where}\:\mathbf{v}_{p}^{n+\nicefrac{1}{2},k}=\frac{\mathbf{u}_{p}^{n+1,k}+\mathbf{u}_{p}^{n}}{\gamma_{p}^{n+1,k}+\gamma_{p}^{n}}, & \mathrm{and} & \bar{\mathbf{v}}_{p}^{n+\nicefrac{1}{2},k}=\frac{\mathbf{u}_{p}^{n+1,k}+\mathbf{u}_{p}^{n}}{2\sqrt{1+\left(\frac{\mathbf{u}_{p}^{n+1,k}+\mathbf{u}_{p}^{n}}{2c}\right)^{2}}}.\\
\mathrm{Increment}\:\:k.
\end{eqnarray*}
 The number of iterations is currently specified to a fixed number.
\item Update timestep counter and return to 1.
\end{enumerate}
\end{algorithm}
The energy- and charge-conserving algorithm has been implemented in
a code named iVPIC, based on the VPIC code developed at Los Alamos
National Laboratory \citep{bowers2008ultrahigh}. Our iterative implementation
is outlined in Algorithm \ref{alg:Iterative-solution}. Only Ampere's
equation needs iteration, as it couples the particles and the electric
field nonlinearly. However, given that we respect the CFL for stability,
the coupling is not stiff numerically, and a simple Picard iteration
is sufficient for fast convergence. In practice, we observe convergence
to numerical round-off (in single precision) in around 5 iterations.
Note that particles are pushed once per iteration, at a cost comparable
to an explicit particle push. Therefore, except for the additional
storage needed to keep old-time particle quantities available during
the iteration, the particle push in our implementation is competitive
with a state-of-the-art explicit push.

\section{Numerical results}

We exercise iVPIC on some simple test problems that demonstrate the
correctness of the implementation and the advertised conservation
properties. In particular, we consider a relativistic two-stream instability,
a Weibel instability, and a stimulated Brillouin scattering (SBS)
of a laser in a plasma. All results are obtained by simulations using
single (32-bit) precision IEEE floating-point rounding arithmetic,
as what is originally employed in the VPIC code \citep{bowers2008ultrahigh}.

\subsection{Two-stream instability}

We consider a periodic relativistic system with two counter-streaming
electron beams in an infinitely massive stationary ion background.
The cold beam equations can be written as
\begin{align*}
\frac{\partial n_{j}}{\partial t}+\nabla\cdot(n_{j}\mathbf{v}_{j}) & =0,\\
\frac{\partial\mathbf{u}_{j}}{\partial t}+(\mathbf{v}_{j}\cdot\nabla)\mathbf{u}_{j} & =\frac{q}{m}(\mathbf{E}+\mathbf{v}_{j}\times\mathbf{B}),
\end{align*}
where the $n_{j}$ and $\mathbf{v}_{j}$ are the density and velocity
of electron beam $j$, and $\mathbf{u}_{j}=\mathbf{v}_{j}\gamma_{j}$.
Let $N_{j}$, $\mathbf{V}_{j}$, and $\tilde{n}_{j}$, $\tilde{\mathbf{v}}_{j}$
be the zeroth- and first-order electron density velocity for the $j^{th}$
beam, respectively. We assume the zeroth-order $\mathbf{E}=\mathbf{B}=0$,
and denote $\tilde{\mathbf{E}}$ the first-order electric field. If
the first-order quantities vary as $e^{i(\mathbf{k}\cdot\mathbf{r}-\omega t)}$,
the linearized equations in the Fourier space may be written as
\begin{align}
-\omega\tilde{n}_{j}+N_{j}\mathbf{k}\cdot\tilde{\mathbf{v}}_{j}+\mathbf{k}\cdot\mathbf{V}_{j}\tilde{n}_{j} & =0,\label{eq:linear continuity}\\
-\omega\tilde{\mathbf{u}}_{j}+\mathbf{k}\cdot\mathbf{V}_{j}\tilde{\mathbf{u}}_{j} & =-i\frac{q}{m}\tilde{\mathbf{E}},\label{eq:linear momentum}
\end{align}
where we have used the electrostatic approximation, i.e., the first-order
magnetic field is zero. Note that the first-order relation between
the proper velocity and velocity is 
\begin{equation}
\mathbf{\tilde{u}}_{j}=\tilde{\mathbf{v}}_{j}\Gamma_{j}^{3},\label{eq:linear u}
\end{equation}
where $\Gamma=(1-V^{2}/c^{2})^{-1/2}$. The linearized Poisson equation
is
\begin{equation}
\epsilon_{0}\mathbf{k}\cdot\mathbf{\tilde{E}}=\sum_{j}q_{j}\tilde{n}_{j}.\label{eq:linear Poisson}
\end{equation}
The dispersion relation of the system is found to be 
\begin{equation}
\sum_{j=1}^{2}\frac{\omega_{bj}}{(\omega-\mathbf{k}\cdot\mathbf{V}_{j})}=1,\label{eq:two-stream dispersion}
\end{equation}
where $\omega_{b}=\omega_{p}/\Gamma^{3}$, and $\omega_{p}=\sqrt{Nq/\epsilon_{0}m}$.
Equation \ref{eq:two-stream dispersion} has exactly the same form
of the non-relativistic two-stream dispersion relation except for
the $1/\Gamma^{3}$ factor. For the simplest case, i.e., two beams
with opposite velocities and equal densities ($V_{1}=-V_{2}=V_{0}$,
$\mathbf{k}\cdot\mathbf{V}=kV_{0}$, $N_{1}=N_{2}=N_{0}$), an analytical
solution for the maximum growth rate exists: 
\begin{equation}
\textrm{Im}[\omega]=\omega_{b}/2,\label{eq:growth-rate-relativistic-two-stream}
\end{equation}
with $kV_{0}/\omega_{b}=\sqrt{3}/2$. 

\begin{figure}
\includegraphics[scale=0.45]{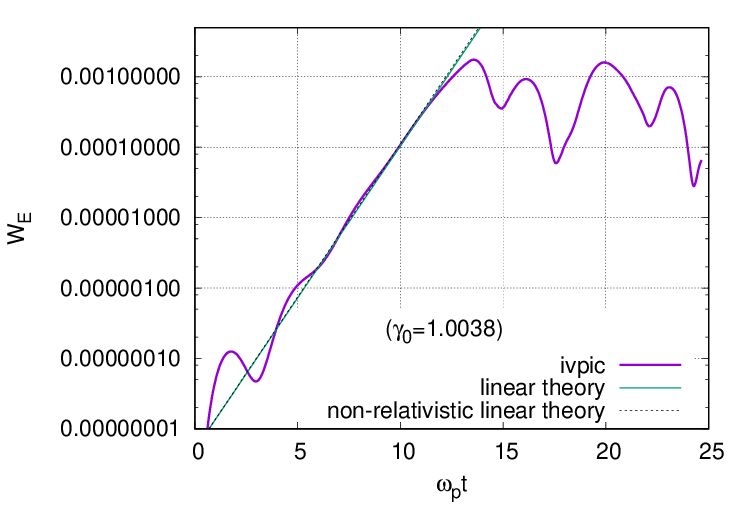}\includegraphics[scale=0.45]{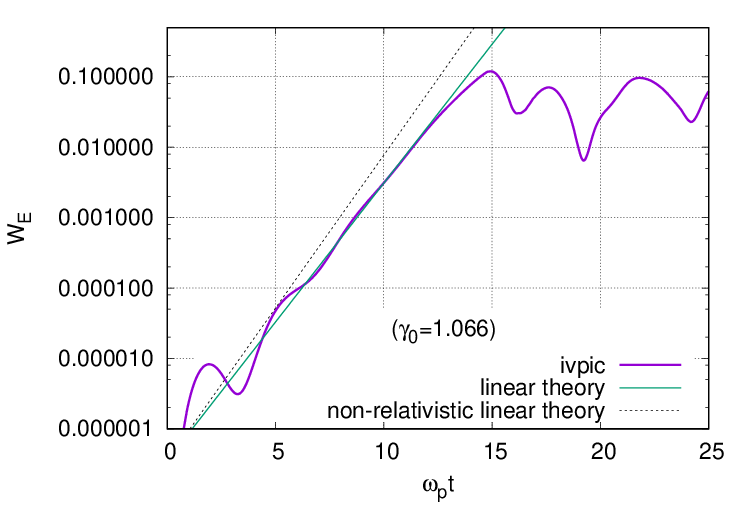}\includegraphics[scale=0.45]{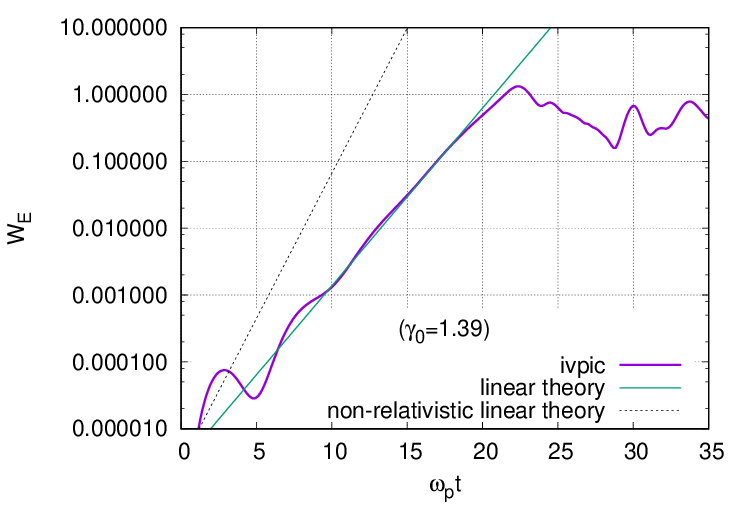}

\caption{\label{fig:Relativistic-two-stream-instabil}Time history of electric
field energy ($W_{E}$) of relativistic two-stream instability simulations
compared with linear theories. }
\end{figure}
We set up simulations with $V_{0}$ varying from non-relativistic
to relativistic regimes.\textcolor{black}{{} The domain size is set
to be $L_{x}=2\pi d_{e}$, where $d_{e}$ is electron skin depth,
with $N_{x}=32$ cells, $N_{pc}=200$ particles per cell. We employ
a time step $\Delta t=0.99\Delta x/c$. The boundary conditions are
periodic. }Figure \ref{fig:Relativistic-two-stream-instabil} shows
the simulation results. We see that, in the non-relativistic regime
($\Gamma\simeq1$), the simulation linear growth rate matches well
with both Eq. \ref{eq:growth-rate-relativistic-two-stream} and the
non-relativistic one, given by: 
\begin{equation}
\textrm{Im}[\omega_{nr}]=\omega_{p}/2.\label{eq:growth-rate-nonrelativistic-two-stream}
\end{equation}
As $V_{0}$ increases, the relativistic lowering of the growth rate
is more noticeable, as expected from the $\Gamma^{-3}$ scaling. For
$\Gamma=1.39$, the relativistic growth rate decreases by a factor
of $\sim0.37$. In both weakly and strongly relativistic simulations,
the change of the growth rate is very well captured. 

\subsection{Weibel instability}

We use the Weibel instability to test the long time-scale conservation
properties of the algorithm. The simulation is performed in a periodic
1D domain of $L_{x}=10d_{e}$ . We employ $N_{x}=64$ cells, $N_{pc}=200$
number of particles per cell, and a time step $\Delta t=0.99\Delta x/c$.
Thermal temperatures of both electrons and ions are set to $v_{th,x}=0.1c$
and $v_{th,y,z}=0.3c$, respectively. The cell size is about 3 Debye
lengths along $x$ at $t=0$. The mass ratio is set to $m_{i}/m_{e}=1836$.
The boundary conditions are periodic.

\begin{figure}
\begin{centering}
\includegraphics{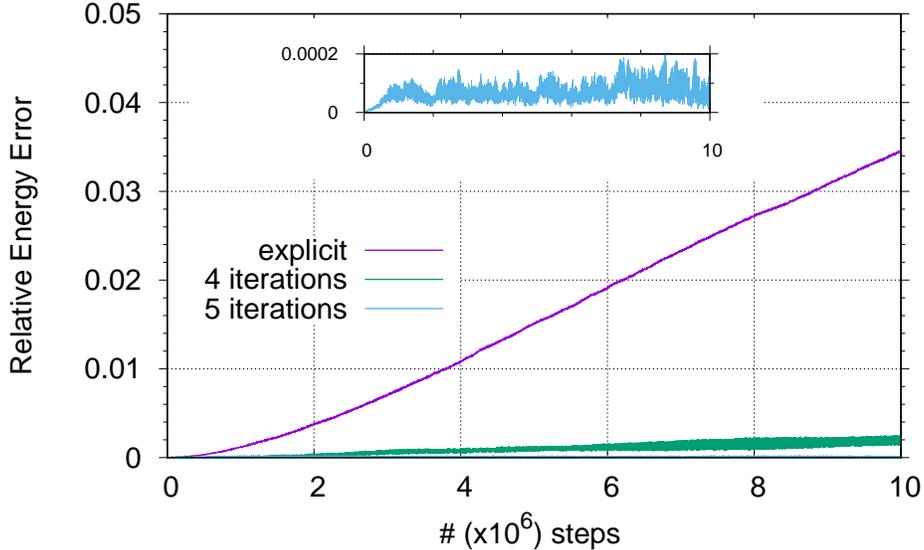}
\par\end{centering}
\caption{\label{fig:History-of-total-energy}History of total energy conservation
errors in long-term electron Weibel instability simulations. Energy
conservation improves with increasing number of iterations. The insert
shows the relative energy error of semi-implicit simulation with 5
iterations.}
\end{figure}
We consider energy conservation first. Fig. \ref{fig:History-of-total-energy}
shows the total energy error history as a function of the number of
timesteps (up to $10^{7}$) for both explicit and semi-implicit simulations
with different number of nonlinear iterations. It is worth noting
that the original VPIC algorithm uses the method of Villasenor and
Buneman \citep{villasenor1992rigorous}, which has been shown to have
much better energy conservation properties than the standard momentum-conserving
schemes \citep{pukhov1999three}. Strictly speaking, the discrete
conservation law proposed in this study ensures conservation of total
energy with the magnetic energy defined by $W_{B}=\frac{1}{2}\int d\mathbf{x}\mathbf{B}^{n+1/2}\cdot\mathbf{B}^{n-1/2}$,
which is not always well-posed (see Appendix\,\ref{sec:Well-posedness-of-magnetic}
for detailed discussion about $W_{B}$). However, $W_{B}$ is almost
always well-posed except in pathological cases, as long as the light
wave mode of interest are well resolved and the CFL condition is respected.
On the other hand, $\hat{W}_{B}=\frac{1}{2}\int d\mathbf{x}\mathbf{B}^{n}\cdot\mathbf{B}^{n}$
at integer timesteps, which can be obtained from Eq. \ref{eq:dBdt-2},
is always positive and well-posed. We have found similar long-term
behavior of both definitions (i.e., conserving one measure leads to
bounded errors in the other). For the semi-implicit scheme, a few
iterations have to be performed to maintain small energy errors. If
a single iteration is used, which is equivalent to a first-order forward
Euler method, the corresponding energy error would grow even faster
than the explicit one (which is second-order accurate). Energy conservation
errors decrease rapidly with the number of iterations. Figure \ref{fig:History-of-total-energy}
shows the relative error of total energy defined at integer timesteps
(employing $\hat{W}_{B}$) up to $10^{7}$ timesteps. For this test,
5 nonlinear iterations are sufficient to maintain conservation errors
at relatively low levels ($\sim10^{-4})$. 

\begin{figure}
\begin{centering}
\includegraphics{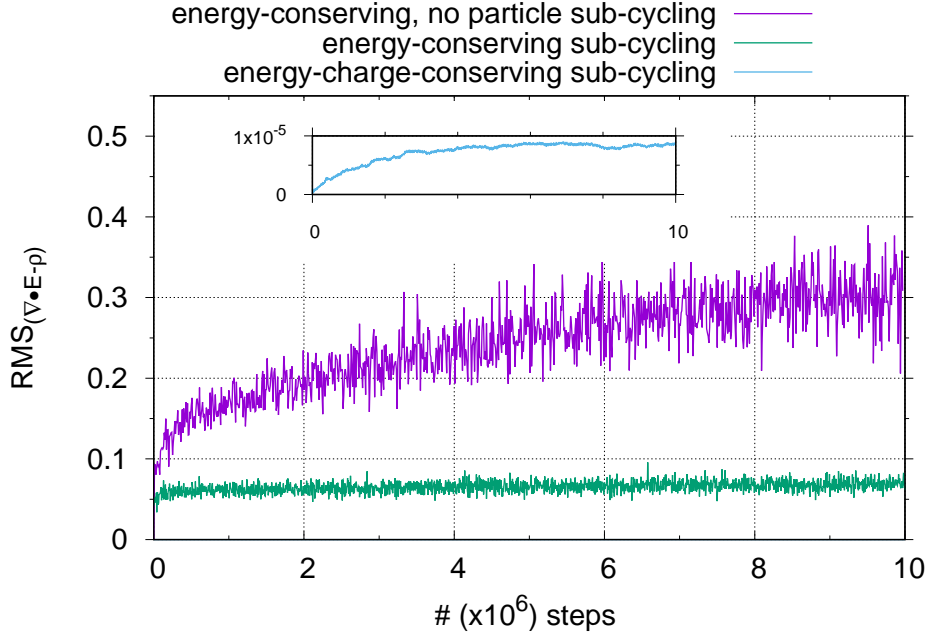}
\par\end{centering}
\caption{\label{fig:History-of-Gauss-law}History of errors of Gauss's law
in long-term electron Weibel instability simulations. The insert shows
the root-mean-square error of Gauss's equation an semi-implicit simulation
using the proposed energy- and charge-conserving algorithm.}
\end{figure}
 Regarding charge conservation, Fig. \ref{fig:History-of-Gauss-law}
shows the root-mean-square of Gauss's law evaluated on the mesh with
different flavors of the semi-implicit solver, all of them energy
conserving, but differing in the particle-push treatment. In particular,
we consider three types of particle pusher: a single particle push
per field step, a subcycled particle push but without special treatment
at cell boundaries (\textcolor{black}{with two subcycling steps}),
and the proposed single-step charge-conserving particle push. As discussed
earlier (Sec. \ref{subsec:Preservation-of-the}), Gauss's law should
be always satisfied everywhere if it is satisfied at $t=0$. Clearly,
for the non-charge-conserving single-step particle-push case, Gauss's
law is violated and the errors accumulate secularly with the number
of time steps. Particle sub-cycling improves the long-term behavior
of error accumulation, but the level of error increases very quickly
at the early stage of the simulation and saturates at relatively high
level. With the energy- and charge- conserving scheme, Gauss's law
is satisfied with high accuracy during the simulation, and errors
remain at a very low level ($\thicksim10^{-5}$) .

\subsection{Laser-plasma instabilities}

\begin{figure}
\begin{centering}
\includegraphics[scale=0.7]{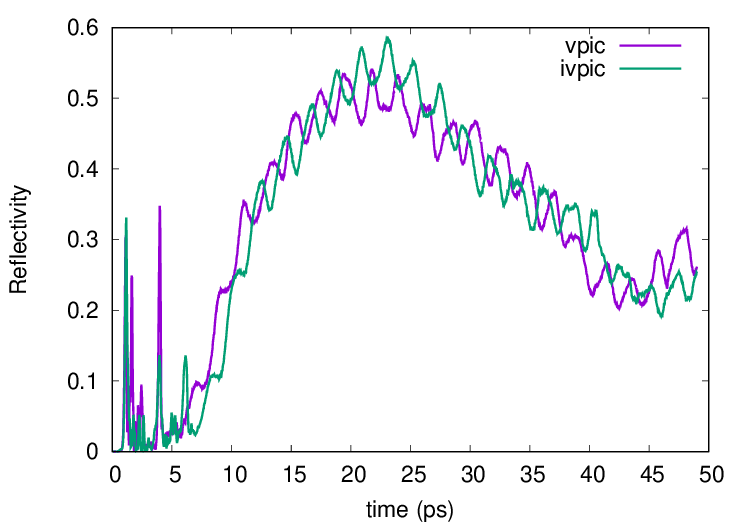}\includegraphics[scale=0.7]{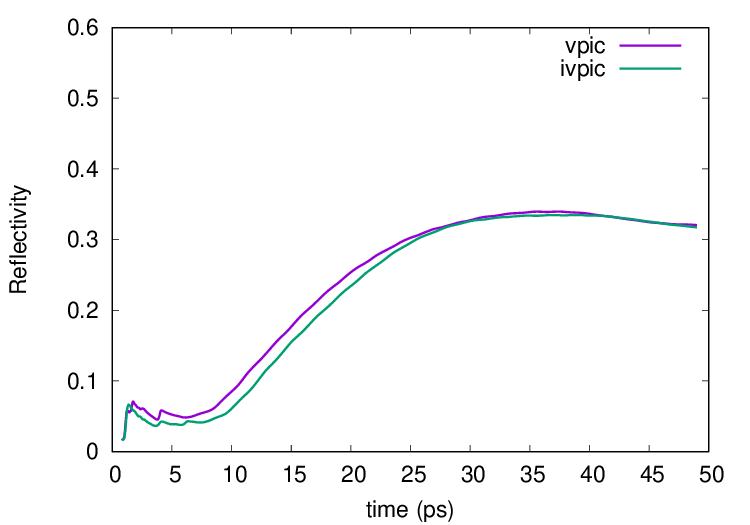}
\par\end{centering}
\caption{\label{fig:History-of-plx}History of reflectivity in  simulations
with both VPIC and iVPIC (using 5 field-particle iterations). Left:
instantaneous reflectivity averaged over laser period. Right: running
time-averaged reflectivity (by $\frac{1}{T}\int_{0}^{T}r(t)dt$).}
\end{figure}

We use the iVPIC code with a plane wave source at intensity $1.25x10^{15}\:\mathrm{W}/\mathrm{cm}^{2}$to
simulate stimulated Raman scattering (SRS) and stimulated Brillouin
scattering (SBS) in a single laser speckle , in a under-dense plasma
($n_{e}=0.05n_{cr}$) where $n_{cr}$ is the critical density, similar
to that of Ref. \citep{albright2016multi}. The simulation is performed
in a quasi-1D domain of size $320\times0.11$ ($d_{e}^{2}$), with
grid points $N_{x}\times N_{z}=7776\times2$, and the number of particles
per cell $N_{pc}=512$. timestep $\Delta t=0.98\,\Delta t_{CFL}$,
where $\Delta t_{CFL}=\Delta x/\sqrt{2}c$ in two dimensions. The
laser is polarized along $\hat{y}$ and the simulation is in the $\hat{x}-\hat{z}$
domain. We initiate the  laser pulse by gradually ramping up the intensity
at the $x=0$ boundary, and the laser propagates into the plasma.
We employ absorbing boundary conditions for fields, and absorbing
boundary conditions in $x$ and periodic in $z$ for particles. The
laser speckle is modeled as a Gaussian laser pulse polarized along
the $y$ direction. The same setup is simulated with both VPIC and
iVPIC. Figure \ref{fig:History-of-plx} shows the backscatter reflectivity
($r=1-$measured Poynting flux/Laser Poynting flux) obtained at $x=0$
from the quasi-1D simulation as a function of time. Except for some
differences in the fluctuations that are sensitive to the thermal
noise and random number seeds used, the instantaneous and running
time-averaged the reflectivities agree well between VPIC and iVPIC.
The two simulations reproduce both the bursty nature of SRS in the
electron trapping regime at early times (t < 7 ps) as well as the
transition to SBS at late times (t > 7 ps). Although differences are
observed in the timing and amplitude of the individual bursts (which
are known to be sensitive to the properties of the noise inherent
to particle simulations), excellent agreement is found in the levels
of time-averaged reflectivity between VPIC and iVPIC as well as the
timing of the transition from SRS to SBS.

\section{Discussion and summary}

We have developed the first low-dispersion, energy- and charge-conserving
relativistic Vlasov-Maxwell PIC algorithm. It is semi-implicit, leap-frogging
electromagnetic fields and time-centering the particle push. The light
wave dispersion errors are the same as the standard explicit PIC \citep{birdsall2004plasma},
and lower than the previously proposed fully implicit scheme \citep{lapenta2011particle}.
It has been known that the Yee scheme conserves the discrete energy
of electromagnetic fields \citep{schuhmann2001conservation,edelvik2004general}.
We have extended such a principle to PIC by ensuring that the energy
exchange between particles and fields is treated correctly. It requires
nonlinear iteration between the electric field update in Ampere's
equation and the particle push, but the iteration is not stiff and
converges very quickly with a simple Picard iterative scheme. To ensure
simultaneous rigorous and automatic charge conservation, we have developed
a novel particle push scheme that not only deals with particle cell
crossings effectively without actually stopping the particle at cell
boundaries, but also is exactly energy-conserving and directly revertible
(i.e. can be solved explicitly \citep{higuera2017structure}). The
superior properties of the scheme have been demonstrated on three
numerical examples of varying complexity and dimensionality, from
electrostatic (two-stream) to fully electromagnetic (Weibel, SBS).
At this point, without vectorization and careful optimization, each
iteration in iVPIC is about 50\% slower than an unvectorized explicit
step, resulting in a slowdown of a factor of 6 when 4 iterations are
taken. Future work will focus on the implementation of higher-order
particle-mesh interpolations, and on the vectorization and optimization
of the algorithm on modern architectures.

\appendix
\begin{appendices}

\part*{Appendix}

\section{Well-posedness of magnetic energy definition\label{sec:Well-posedness-of-magnetic}}

We address the question of whether the magnetic energy definition
in Eq. \ref{eq:magnetic energy} is well-posed or not, i.e., whether
the energy is positive and provide a physically meaningful measure.
We show that it is almost always well-posed when the CFL timestep
is respected, and it is a second-order accurate approximation of the
magnetic energy defined at integer time levels. We begin by expressing
$\mathbf{B}^{n\pm1/2}$ around time $n$ as:
\begin{equation}
\mathbf{B}^{n\pm1/2}=\mathbf{B}^{n}\pm\frac{\Delta t}{2}\partial_{t}\mathbf{B}^{n},
\end{equation}
to find:
\begin{equation}
W_{B}=\hat{W}_{B}-\frac{\Delta t^{2}}{8}\int d\mathbf{x}(\partial_{t}\mathbf{B}^{n})\cdot(\partial_{t}\mathbf{B}^{n}),
\end{equation}
where we denote $W_{B}=\frac{1}{2}\int d\mathbf{x}\mathbf{B}^{n+1/2}\cdot\mathbf{B}^{n-1/2}$,
and $\hat{W}_{B}=\frac{1}{2}\int d\mathbf{x}(B^{n})^{2}$. This proves
second-order accuracy. However, \emph{a priori}, the measure $W_{B}$
may not be positive, while $\hat{W}_{B}$ is. For simplicity, we first
show that $W_{B}$ is almost always well-posed in a 1D analysis. 

Assuming that the magnetic field varies as $e^{-i\omega t+ikx}$,
where $\omega=\omega(k)$,the discrete Fourier transform may be written
as
\begin{equation}
\mathbf{B}(x,t)=\frac{1}{N_{g}}\sum_{l=0}^{N_{g}-1}\mathbf{B}(\omega_{l},k_{l})e^{-i\omega_{l}t+ik_{l}x},
\end{equation}
where $N_{g}$ is the number of grid points and $k_{l}=\frac{2\pi l}{L}$.
Then, by applying the Fourier transform to $W_{B}$,
\begin{align}
\sum_{h=0}^{N_{g}-1}\mathbf{B}^{n+\nicefrac{1}{2}}(x_{h})\cdot\mathbf{B}^{n-\nicefrac{1}{2}}(x_{h})^{*} & =\frac{1}{N_{g}^{2}}\sum_{h=0}^{N_{g}-1}\sum_{l=0}^{N_{g}-1}\sum_{m=0}^{N_{g}-1}\mathbf{B}(\omega_{l},k_{l})\cdot\mathbf{B}(\omega_{k},k_{m})^{*}e^{i(k_{l}-k_{m})x_{h}}\Re\left[e^{-i(\omega_{l}t^{n+\nicefrac{1}{2}}-\omega_{m}t^{n-\nicefrac{1}{2}})}\right]\nonumber \\
 & =\frac{1}{N_{g}^{2}}\sum_{l=0}^{N_{g}-1}\sum_{m=0}^{N_{g}-1}\mathbf{B}(\omega_{l},k_{l})\cdot\mathbf{B}(\omega_{k},k_{m})^{*}\Re\left[e^{-i(\omega_{l}t^{n+\nicefrac{1}{2}}-\omega_{m}t^{n-\nicefrac{1}{2}})}\right]\sum_{h=0}^{N_{g}-1}e^{i(k_{l}-k_{m})x_{h}}\nonumber \\
 & =\frac{1}{N_{g}}\sum_{l=0}^{N_{g}-1}\mathbf{B}(\omega_{l},k_{l})\cdot\mathbf{B}(\omega_{l},k_{l})^{*}\cos(\omega_{l}\Delta t),
\end{align}
where the superscript $*$ denotes complex conjugate, and $\Re[]$
denotes the real part of a complex number. We have used the orthogonal
property of Fourier modes:
\begin{equation}
\sum_{h=0}^{N_{g}-1}e^{i(k_{l}-k_{m})x_{h}}=N_{g}\delta_{lm}.
\end{equation}
From the 1D light wave dispersion relation:
\begin{equation}
\left[\frac{\Delta_{h}}{c\Delta t}\right]^{2}\sin^{2}\left(\frac{\omega\Delta t}{2}\right)=\sin^{2}\left(\frac{k_{x}\Delta_{h}}{2}\right),
\end{equation}
and using that $2\sin^{2}(\theta/2)=1-\cos(\theta)$, we find that:
\begin{equation}
\cos(\omega\Delta t)\geq\left[\frac{c\Delta t}{\Delta_{h}}\right]^{2}\text{\ensuremath{\cos}(\ensuremath{k_{x}\Delta_{h}}),}
\end{equation}
when $\frac{c\Delta t}{\Delta_{h}}\leq1$ . Therefore:
\begin{equation}
\sum_{h=0}^{N_{g}-1}\mathbf{B}^{n+\nicefrac{1}{2}}(x_{h})\cdot\mathbf{B}^{n-\nicefrac{1}{2}}(x_{h})^{*}\geq\frac{1}{N_{g}}\left[\frac{c\Delta t}{\Delta_{h}}\right]^{2}\sum_{l=0}^{N_{g}-1}\mathbf{B}(\omega_{l},k_{l})\cdot\mathbf{B}(\omega_{l},k_{l})^{*}\text{\ensuremath{\cos}(\ensuremath{k_{l}\Delta_{h}}).}
\end{equation}
For the case of white noise (equal spectral content in all frequencies),
it is easy to see that:
\begin{equation}
\frac{1}{N_{g}}\sum_{l=0}^{N_{g}-1}\mathbf{B}(\omega_{l},k_{l})\cdot\mathbf{B}(\omega_{l},k_{l})^{*}\cos(k_{l}\Delta_{h})=A\sum_{l=0}^{N_{g}-1}\cos(k_{l}\Delta_{h})=0,
\end{equation}
where $A$ is a positive constant, and provided that $N_{g}$ is an
even number. It follows that:
\begin{equation}
\sum_{h=0}^{N_{g}-1}\mathbf{B}^{n+\nicefrac{1}{2}}(x_{h})\cdot\mathbf{B}^{n-\nicefrac{1}{2}}(x_{h})^{*}\geq0,
\end{equation}
i.e., the magnetic energy is positive definite if the CFL constraint
$\frac{\Delta tc}{\Delta_{h}}<1$ is respected. In general, simulations
will have varying spectral content with frequency, but low-frequency,
well-resolved modes will carry the most information, and therefore
the white noise analysis is conservative. For a mode with more than
four grid points per wavelength, $\cos(k_{l}\Delta_{h})>0$. Therefore,
as long as the field energy is not dominantly associated with high
$k$ (i.e. $\left|k\Delta_{h}\right|>\pi/2$) modes, the field energy
defined from $\mathbf{B}^{n+1/2}\cdot\mathbf{B}^{n-1/2}$ will be
positive.

The above analysis can be straightforwardly extended to multiple dimensions.
Assume a uniform grid, i.e., $\Delta x=\Delta y=\Delta z=\Delta_{h}$,
and $\frac{\Delta tc}{\Delta_{h}}\lesssim\frac{1}{\sqrt{d}}$, where
$d$ is the number of dimensions. Using the numerical dispersion relation
of the light wave in 3D, i.e., 
\begin{equation}
\left[\frac{\Delta_{h}}{c\Delta t}\right]^{2}\sin^{2}\left(\frac{\omega\Delta t}{2}\right)=\sin^{2}\left(\frac{k_{x}\Delta_{h}}{2}\right)+\sin^{2}\left(\frac{k_{y}\Delta_{h}}{2}\right)+\sin^{2}\left(\frac{k_{z}\Delta_{h}}{2}\right),
\end{equation}
we find that:
\begin{equation}
\cos(\omega\Delta t)\geq\left[\frac{c\Delta t}{\Delta_{h}}\right]^{2}\text{\ensuremath{\left[\cos(k_{x}\Delta_{h})+\cos(k_{y}\Delta_{h})+\cos(k_{z}\Delta_{h})\right]}.}
\end{equation}
Clearly, the field energy given by Eq. \ref{eq:magnetic energy} is
well-posed if it is well-posed in each direction.

\section{Equivalence of Villasenor and Buneman's charge conservation scheme
and the proposed one\label{sec:Equvilence-to-Villasenor}}

The proposed current and charge density deposition scheme and cell-crossing
scheme are exactly the same as Villasenor and Buneman's \citep{villasenor1992rigorous}.
The derivation presented in Sec. \ref{subsec:Charge-conservation}
appears to be new, however, and can be easily extended to higher-order
splines \citep{stanier2019fully}. Here we explicitly demonstrate
the equivalence of the two schemes.

Assume that the particle trajectory is from $(x_{p}^{n},y_{p}^{n},z_{p}^{n})$
to $(x_{p}^{n+1},y_{p}^{n+1},z_{p}^{n+1})$, which lies within a single
cell. We employed B-splines which are written as (letting $\Delta x=\Delta y=\Delta z=1$,
and $x_{i}=y_{j}=z_{k}=0$, without loss of generality): 
\begin{align}
S_{0}(x_{i+\nicefrac{1}{2}}-x_{p}) & =1,\label{eq:NGP}\\
S_{1}(x_{i}-x_{p}) & =1-x_{p},\label{eq:Tent1}\\
S_{1}(x_{i+1}-x_{p}) & =x_{p},\label{eq:Tent2}
\end{align}
and similarly in $y$ and $z$ directions.

We first show that Eq. \ref{eq:cc-1d} is trivally satisfied. Assuming
that the particle is in a cell centered at ($i+\nicefrac{1}{2}$),
we write the equation for the node ($i$) and $(i+1)$ respectively
as 
\begin{align}
\mathrm{lhs} & =\frac{(1-x_{p}^{n+1})-(1-x_{p}^{n})}{\Delta t}+v_{x}^{n+\nicefrac{1}{2}}\frac{1-0}{1}=0,\\
\mathrm{lhs} & =\frac{x_{p}^{n+1}-x_{p}^{n}}{\Delta t}+v_{x}^{n+\nicefrac{1}{2}}\frac{0-1}{1}=0,
\end{align}
where we have used Eq. \ref{eq:CN-x}. A formal derivation can be
found in Ref. \citep{chen2011energy}.

Adopting similar notations as in Ref. \citep{villasenor1992rigorous},
we denote $\Delta x_{p}=x_{p}^{n+1}-x_{p}^{n}$, $\Delta y_{p}=y_{p}^{n+1}-y_{p}^{n}$,
$\Delta z_{p}=z_{p}^{n+1}-z_{p}^{n}$, and $\bar{\xi}=(x_{p}^{n+1}+x_{p}^{n})/2$,
$\bar{\eta}=(y_{p}^{n+1}+y_{p}^{n})/2$, $\bar{\zeta}=(z_{p}^{n+1}+z_{p}^{n})/2$,
Eqs. \ref{eq:rhon}-\ref{eq:jz} for the node $(i+1,j+1,k+1)$ are
written as
\begin{align}
\rho_{p,i+1,j+1,k+1}^{n+1} & =q_{p}(\bar{\xi}+\frac{1}{2}\Delta x_{p})(\bar{\eta}+\frac{1}{2}\Delta y_{p})(\bar{\zeta}+\frac{1}{2}\Delta z_{p}),\\
\rho_{p,i+1,j+1,k+1}^{n} & =q_{p}(\bar{\xi}-\frac{1}{2}\Delta x_{p})(\bar{\eta}-\frac{1}{2}\Delta y_{p})(\bar{\zeta}-\frac{1}{2}\Delta z_{p}),\\
j_{x,p,i+\nicefrac{1}{2},j+1,k+1}^{n+\nicefrac{1}{2}} & =q_{p}(\frac{\Delta x_{p}}{\Delta t}\bar{\eta}\bar{\zeta}+\frac{1}{12}\Delta y_{p}\Delta z_{p}),\\
j_{y,p,i+1,j+\nicefrac{1}{2},k+1}^{n+\nicefrac{1}{2}} & =q_{p}(\frac{\Delta y_{p}}{\Delta t}\bar{\zeta}\bar{\xi}+\frac{1}{12}\Delta z_{p}\Delta x_{p}),\\
j_{z,p,i+1,j+1,k+\nicefrac{1}{2}}^{n+\nicefrac{1}{2}} & =q_{p}(\frac{\Delta z_{p}}{\Delta t}\bar{\xi}\bar{\eta}+\frac{1}{12}\Delta x_{p}\Delta y_{p}),\\
j_{x,p,i+\nicefrac{3}{2},j+1,k+1}^{n+\nicefrac{1}{2}} & =0,\\
j_{y,p,i+1,j+\nicefrac{3}{2},k+1}^{n+\nicefrac{1}{2}} & =0,\\
j_{z,p,i+1,j+1,k+\nicefrac{3}{2}}^{n+\nicefrac{1}{2}} & =0,
\end{align}
and their substitutions into Eq. \ref{eq:disc-cc} yield 
\begin{equation}
\Delta x_{p}\bar{\eta}\bar{\zeta}+\Delta y_{p}\bar{\zeta}\bar{\xi}+\Delta z_{p}\bar{\xi}\bar{\eta}=(\bar{\xi}+\frac{1}{2}\Delta x_{p})(\bar{\eta}+\frac{1}{2}\Delta y_{p})(\bar{\zeta}+\frac{1}{2}\Delta z_{p})-(\bar{\xi}-\frac{1}{2}\Delta x_{p})(\bar{\eta}-\frac{1}{2}\Delta y_{p})(\bar{\zeta}-\frac{1}{2}\Delta z_{p}),
\end{equation}
which is exactly the same as Eq. 38 of Ref. \citep{villasenor1992rigorous}
except for the subscript $p$ adopted in this study.

For trajectories across cell boundaries, the treatment is exactly
the same as Villasenor and Buneman's scheme, i.e., we split the trajectory
into segments, each of which lies in one cell, and the above treatment
applies.

\end{appendices}

\pagebreak\bibliographystyle{ieeetr}
\bibliography{ivpic}

\end{document}